\documentclass[prd,twocolumn,showpacs,preprintnumbers,amsmath,amsfonts,amssymb,nofootinbib]{revtex4}
\usepackage[latin1]{inputenc}
\usepackage{lmodern} 
\usepackage{graphicx}

\usepackage{amsmath}
\usepackage{amsfonts}
\usepackage{amssymb}
\usepackage{subfigure}
\usepackage[breaklinks]{hyperref}

\newcommand{\tr}{{\rm Tr\,}}
\newcommand{\nn}{{\nonumber}}
\newcommand{\Li}{{\text{Li}_2}}

\begin{document}

\title{Equivalence of Wilson Loops in ABJM and $\mathcal{N}=4$ SYM Theory}
\date{October 7, 2011}
\author{Konstantin Wiegandt}
\thanks{\href{mailto:konstantin.wiegandt@physik.hu-berlin.de}{konstantin.wiegandt@physik.hu-berlin.de}}
\affiliation{Institut f\"ur Physik, Humboldt-Universit\"at zu Berlin, Newtonstra\ss e 15, D-12489 Berlin, Germany}
\preprint{HU-EP-11/48}

\begin{abstract}
In previous investigations, it was found that four-sided polygonal light-like Wilson loops in ABJM theory calculated to two-loop order have the same form as the corresponding Wilson loop in $\mathcal{N} = 4$ SYM at one-loop order. Here we study light-like polygonal Wilson loops with $n$ cusps in planar three-dimensional Chern-Simons and ABJM theory to two loops. Remarkably, the result in ABJM theory precisely agrees with the corresponding Wilson loop in $\mathcal{N} = 4$ SYM at one-loop order for arbitrary $n$. In particular, anomalous conformal Ward identites allow for a so-called remainder function of conformal cross ratios for $n \geq 6$, which is found to be trivial at two loops in ABJM theory in the same way as it is trivial in $\mathcal{N} = 4$ SYM at one-loop order. Furthermore, the result for arbitrary $n$ obtained here, allows for a further investigation of a Wilson loop / amplitude duality in ABJM theory, for which non-trivial evidence was recently found by a calculation of four-point amplitudes that match the Wilson loop in ABJM theory.
\end{abstract}

\maketitle

 \section{Introduction}
Our motivation to consider polygonal light-like Wilson loops in 3d Chern-Simons and $\mathcal{N}=6$ superconformal Chern-Simons (ABJM) theory \cite{Aharony:2008ug} stems from the Wilson loop/scattering amplitude duality in $\mathcal{N} = 4$ super Yang-Mills.

In planar $\mathcal{N}=4$ super Yang-Mills $n$-particle  MHV scattering amplitudes $\mathcal{A}_n^{\text{MHV}}=A_n^{\text{tree}}M_n$ are related to the 
expectation value of the $n$-cusped Wilson loop operator 
\begin{equation}\label{eqn:Wilson-loop-operator}
 \langle W_{n} \rangle :=  \frac{1}{N}
\, \langle 0 |\,
\tr \mathcal{P} \exp \left( i \oint_{\mathcal{C}_n} A_\mu dz^\mu \right)\, | 0\rangle\,.
\end{equation}
The contour of the $n$-sided polygon $\mathcal{C}_n$ is given by $n$ points $x_i$ $(i=1,...,n)$ which are related to the massless particle momenta via $x_{i+1}-x_i=p_i$. The segments of the contour are thus light-like, i.e. $(x_i-x_{i+1})^2=p_i^2=0$. 

The relation between the Wilson loop and the scattering amplitude is given by 

\begin{equation}\label{eqn:duality-relation}
 \ln M_n = \ln \langle W_n \rangle + \text{const}\,.
\end{equation}

This duality was discovered in the dual $AdS_{5}\times S^{5}$ string picture at 
strong gauge coupling in \cite{Alday:2007hr} 
and shown to exist also in the weak coupling regime \cite{Drummond:2007aua,Brandhuber:2007yx,Drummond:2007cf}
with profound consequences on the symmetries
of these correlators leading to a dual superconformal \cite{Drummond:2008vq} 
respectively Yangian symmetry \cite{Drummond:2009fd}
of scattering amplitudes, for reviews see \cite{Alday:2008yw,Henn:2009bd}. For a rieview of AdS/CFT integrability see \cite{Beisert:2010jr}.

It was found that the expectation value of the Wilson loop in $\mathcal{N}=4$ super Yang-Mills  is governed by an anomalous conformal Ward identity that completely fixes its form at 4 and 5 points and allows for an arbitrary function of conformal invariants starting from 6 points. This so-called remainder function $R_n$ is indeed present starting from 6 points and leads to a correction \cite{Drummond:2007bm}, \cite{Bern:2008ap} of the BDS ansatz \cite{Bern:2005iz} for planar gluon scattering amplitudes.

Recently, the duality has been extended to amplitudes with arbitrary helicity states by introducing a suitable supersymmetric Wilson loop \cite{CaronHuot:2010ek,Mason:2010yk}.

Furthermore, a duality between light-like Wilson loops with $n$ cusps and $n$-point correlation functions of half-BPS protected operators in the limit where the positions of adjacent operators become light-like separated was established in \cite{Alday:2010zy,Eden:2010ce,Eden:2011yp,Eden:2011ku,Adamo:2011dq}.

From the string perspective the scattering amplitude/Wilson loop duality in the
$AdS_{5}/CFT_{4}$ system arises from
a combination of bosonic and fermionic T-dualities under which the free 
$AdS_{5}\times S^{5}$ superstring is self-dual \cite{Berkovits:2008ic,Beisert:2008iq}.
Hence, for the existence of an analogue duality in ABJM theory one would 
require a similar self-duality of the $AdS_{4}\times \mathbb{CP}_{3}$ superstring under the
combined T-dualities. The problem was analysed in \cite{Grassi:2009yj,Adam:2010hh,Adam:2009kt,Dekel:2011qw,Bakhmatov:2010fp,Bakhmatov:2011aa} but no T-self-duality could be established so far. 

At tree-level, recent developments have uncovered Yangian and dual superconformal symmetry of the amplitudes in ABJM theory \cite{Bargheer:2010hn,Lee:2010du,Huang:2010qy,Gang:2010gy,Lipstein:2011ej}. In \cite{Agarwal:2008pu}  a vanishing result for the four-point one-loop amplitudes in ABJM theory was found and the authors speculated whether the two-loop scattering amplitudes in $\mathcal{N} = 6$ Chern-Simons could be simply related to the one-loop $\mathcal{N} = 4$ Yang-Mills amplitudes.

In \cite{Henn:2010ps} we calculated the expectation value of the  Wilson loop operator \eqref{eqn:Wilson-loop-operator} in the planar limit for light-like polygonal contours $\mathcal{C}_n$ in pure Chern-Simons and ABJM theory. 

Conformal Ward identities force $\langle W_{n}\rangle_{\text{1-loop}}$ 
to depend only on conformally invariant cross ratios of the $(x_i-x_j)^{2}$.  At one-loop order in pure Chern-Simons and ABJM theory we found that the correlators with four and six cusps vanish, leading to the conclusion that the allowed function of conformal cross ratios is trivial at six points. We thus conjectured the n-point correlator to vanish
\begin{equation}
\langle W_n \rangle^{\text{CS}}_{\text{1-loop}} = 0
\end{equation}
which was indeed proven in \cite{Bianchi:2011rn} and also non-trivial evidence for a duality between Wilson loops and correlators in ABJM theory was found  at one-loop level.

Furthermore, we computed the tetragonal Wilson loop $W_{4}$  at two-loop order in pure Chern-Simons and  ABJM theory. Remarkably, the result in dimensional
reduction regularisation with $d=3-2\epsilon$ for the correlator in  ABJM theory is
of the same functional form as the one-loop result in $\mathcal{N}=4$ super
Yang-Mills theory.
Most interestingly, it was recently found by two independent approaches, using generalized unitarity methods in \cite{Chen:2011vv} and by a direct superspace Feynman diagram calculation in \cite{Bianchi:2011dg}, that the two-loop result for four-point scattering amplitudes in ABJM theory agrees with the Wilson loop 
\begin{align}
M_4^{(2)} = \frac{A_4^{(2)}}{A_4^{\text{tree}}} = \langle W_4 \rangle^{\text{ABJM}}_{\text{2-loop}}+ const. 
\end{align}
upon a specific identification of the regularisation scales \cite{Bianchi:2011dg,Chen:2011vv}.
This establishes the first non-trivial example for a Wilson loop / amplitude duality of the form \eqref{eqn:duality-relation} in ABJM  theory. Very recently, these result were extended to the more general case of ABJ theory in \cite{Bianchi:2011fc}.

In light of these recent findings on structural similarities between observables in $\mathcal{N}=4$ super Yang-Mills and ABJM theory, it is natural to ask, whether the duality between Wilson loops and amplitudes in ABJM theory continues to hold beyond $n=4$, as it does in $\mathcal{N}=4$.  
  
In this work we perform numerical computations to extend our findings of \cite{Henn:2010ps} to the $n$-sided Wilson loop at two-loop order. Remarkably, we find that the hexagonal Wilson loop at two loops agrees with the corresponding Wilson loop in $\mathcal{N}=4$ super Yang-Mills.

We perform a detailed numerical analysis for the hexagonal Wilson loop leading to a guess for the $n$-point case, which we numerically check also for  $n>6$ in a limited set of kinematical points. Again we find, that the result agrees with the result for the Wilson loop in $\mathcal{N}=4$ super Yang-Mills. It is thus natural to expect the result to hold for all $n$
\begin{align}
\langle W_n \rangle^{\text{ABJM}}_{\text{2-loop}} &=  \left( \frac{N}{k}\right)^2 \Big [-\frac{1}{2}\sum_{i=1}^n  \frac{(-{\mu^\prime}^2 \, x_{i,i+2}^2 )^{2\epsilon}}{(2\epsilon)^2}  \\ \nonumber
& \qquad  \qquad \qquad  + \mathcal{F}_n^{\text{WL}} + r_n
\Big ] 
\end{align}
where ${\mu^{\prime}}^2= \mu^2 8 \pi e^{\gamma_E}$, $r_n$ is a constant that depends linearly on $n$ and is specified below \eqref{eqn:ABJM-result-Wn} and where the finite contribution $\mathcal{F}_n^{\text{WL}}$ is given by the finite part of the Wilson loop in $\mathcal{N}=4$ SYM, which up to a constant\footnote{$\mathcal{F}_n^{\text{WL}}=\mathcal{F}_n^{\text{BDS}}-n/4 \zeta(2)$.} is the finite part in the BDS conjecture \cite{Bern:2005iz}, i.e. for $n=4$ and $n=6$
\begin{align}\label{eqn:BDS-WL-n4-n6}
\mathcal{F}_4^{\text{WL}} &= \frac{1}{2} \ln^2 \left( \frac{x_{13}^2}{x_{24}^2} \right) + \frac{\pi^2}{2} \\ \nn
\mathcal{F}_6^{\text{WL}} &= \frac{\pi^2}{2} +
 \frac{1}{2} \sum_{i=1}^6\Bigg[ - \ln \left(\frac{x^2_{i,i+2}}{x^2_{i,i+3}} \right) \ln \left(\frac{x^2_{i+1,i+3}}{x^2_{i,i+3}} \right) \\ \nn
+& \frac{1}{4} \ln^2 \left( \frac{x^2_{i,i+3}}{x^2_{i+1,i+4}}\right) 
-\frac{1}{2} \text{Li}_2 \left( 1-\frac{x_{i,i+2}^2 x_{i+3,i+5}^2}{ x_{i,i+3}^2 x_{i+2,i+5}^2} \right) \Bigg]\,.
\end{align}

Thus, the Wilson loop in ABJM theory at two-loop order precisely agrees with the form of the Wilson loop or, via the amplitude / Wilson loop duality, with the $n$-point MHV amplitudes at one loop in $\mathcal{N}=4$ super Yang-Mills. 

It would be  very interesting to establish a six-point amplitude calculation in order to see whether the duality relation \eqref{eqn:duality-relation}  in ABJM  theory also holds at six points. Furthermore, it would be interesting to perform a four-point amplitude or Wilson loop computation at four loops, to check whether the conjectured BDS-like ansatz \cite{Bianchi:2011dg} for the four-point amplitude in ABJM indeed holds.

The relations between Wilson loops, amplitudes and correlators seem to hold not only in the special case of the  maximally supersymmetric $\mathcal{N}=4$ SYM theory but also in ABJM theory. 
Since the duality may thus not just be a particular feature of $\mathcal{N}=4$ SYM, it remains an important task to understand the precise origin of the similarity of these different observables in quantum field theory.

\section{N-sided Wilson Loops in CS theory}
The solution of the Ward identity \cite{Henn:2010ps} for the light-like polygonal Wilson loop in pure Chern-Simons theory reads
\begin{align}\label{eqn:solution-special-conformal-ward-identity-xij}\nn
\langle W_n \rangle^{\text{CS}}_{\text{2-loop}} =  -\frac{1}{4}\left(\frac{N}{k}\right)^2  \Big[& 2\ln(2)\sum_{i=1}^n\frac{(-x_{i,i+2}^2 \tilde \mu^2 )^{2\epsilon}}{2\epsilon}\\ 
& + g_{n}(u_{abcd}) + \mathcal{O}(\epsilon) \Big]\,.
\end{align}
where $\tilde{\mu}^2=\mu^2  \pi e^{\gamma_E}$ and $g_n$ is a function that depends only on conformally invariant cross ratios $u_{abcd}=(x_{ab}^2x_{cd}^2)/(x_{ad}^2x_{cb}^2)$ which can be constructed starting from $n=6$.\\ 

\begin{figure}[t]
\centering
\subfigure[]{\begin{minipage}{3.5cm}\centering
		\includegraphics[width=.9 \textwidth]{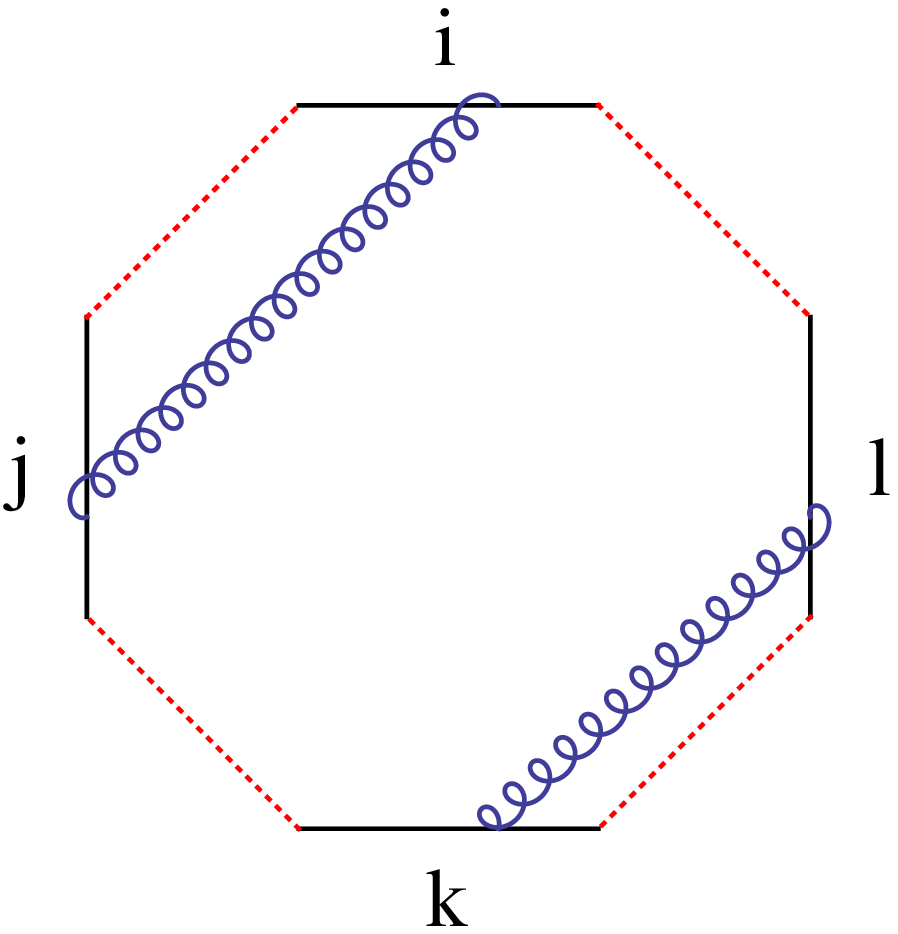}\label{fig: rect1331}~~\\
		\end{minipage}
}
\subfigure[]{\begin{minipage}{3.5cm}\phantom{a}~\\
\centering
		\includegraphics[width=.85 \textwidth]{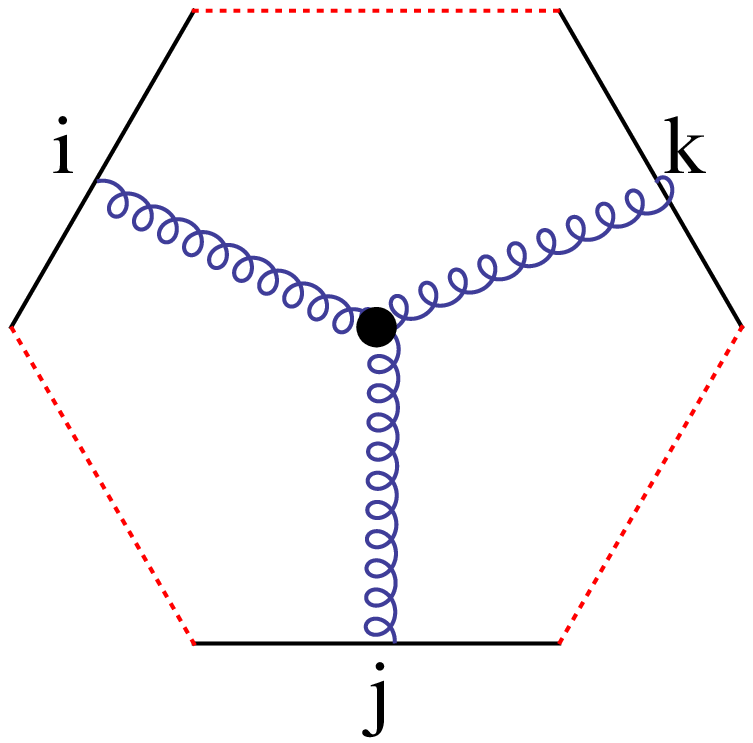}\label{fig:vertexI123}
		\end{minipage}
}
\caption{\ref{fig: rect1331} Two-gluon diagrams $I_{ijkl}$ given in \eqref{eqn:two-gluon-expr},  \ref{fig:vertexI123} vertex diagrams $I_{ijk}$ given in \eqref{eqn:vertexIaandIb}. Dashed lines represent one or more light-like distances.}
\label{fig:2typesofdiagrams}
\end{figure}

At two loops there are two types of contributions to the Wilson loop in pure Chern-Simons theory, one from two-gluon diagrams $\langle W_n\rangle^{\text{two-gluon}}=\sum  I_{ijkl}$ and another one from diagrams involving a three-gluon vertex $\langle W_n\rangle^{\text{vertex}} = \sum I_{ijk}$
\begin{equation}
\langle W_n \rangle^{\text{CS}}_{\text{2-loop}} = \langle W_n\rangle^{\text{two-gluon}} + \langle W_n\rangle^{\text{vertex}} \,.
\end{equation}
The indices $i,j,k,l$ denote the edges that the propagators attach to, see figure \ref{fig:2typesofdiagrams} and their expressions are given in appendix \ref{app:two-gluon}, \ref{app:vertex}. Contributions from gauge- and ghost-loops cancel in dimensional reduction regularisation \cite{Chen:1992ee}, for more details see \cite{Henn:2010ps}.

As explained in \cite{Henn:2010ps}, the vertex diagrams are divergent in the region of integration where all three propagators approach the same edge (all diagrams with more than one propagator on the same edge vanish identically due to the antisymmetry of the Levi-Civita symbol), and we split them up as in \eqref{eqn:split-up-fin-div} 
\begin{equation}
 \langle W_n \rangle^{\text{vertex}} =  \langle W_n \rangle^{\text{div}} + \langle W_n \rangle^{\text{finite}}\,.
\end{equation}
The divergent part can be calculated analytically 
\begin{equation}\nn
\langle W_n \rangle^{\text{div}} = - \left( \frac{N}{k} \right)^2 \left( \frac{\ln (2)}{2}  \sum_{i=1}^n \frac{(-x_{i,i+2}^2 \tilde{\mu}^2)^{2\epsilon}}{2 \epsilon} \right)\,.
\end{equation}
Thus, the function $g_n$ in \eqref{eqn:solution-special-conformal-ward-identity-xij} is given by
\begin{equation}\label{eqn:contributions-to-gn}
 g_n(u_{abcd}) = \langle W_n\rangle^{\text{two-gluon}} + \langle W_n \rangle^{\text{finite}}\,.
\end{equation}
We evaluate these contributions using a Mathematica program that generates all $n$-point diagrams, performs the index-contractions and numerically integrates the diagrams for randomly generated kinematical configurations, for more details see appendix \ref{app:two-gluon-diagrams}.\\~\\
{\bf Hexagonal Wilson loop}\\~\\
For the hexagonal Wilson loop we evaluated the contributions in \eqref{eqn:contributions-to-gn} for a large set of conformally equivalent and conformally non-equivalent kinematical configurations. 

Conformally equivalent configurations must yield the same result, since, by the anomalous conformal Ward identity, the expectation value is constrained  to the form \eqref{eqn:solution-special-conformal-ward-identity-xij}, and thus the function $g_n$ depends only on conformally invariant quantities.

\begin{figure}[t]
\centering
 \includegraphics[width=1 \columnwidth]{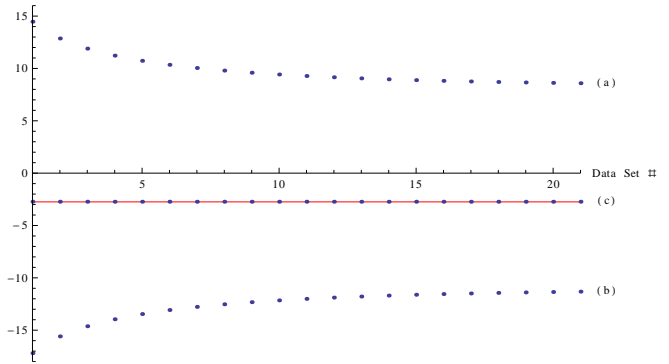}
\caption{Numerical results for the hexagon $n=6$:  $(a)=\langle W_n\rangle^{\text{two-gluon}}$, $(b)=\langle W_n \rangle^{\text{finite}}$ and their sum  $(c) = g_n(u_{abcd})$, see \eqref{eqn:contributions-to-gn}, for the kinematical points given in \ref{app:generation-kinematics}. We omit the common factor $\frac{1}{4}\left(\frac{N}{k}\right)^2$.}
\label{fig:Wvertextwogluonplot}
\end{figure}

It turns out, that even for kinematical configurations which are not conformally equivalent, the unknown function yields the same constant\footnote{The analytical term with $\ln(2)$ in \eqref{eqn:g6},\eqref{eqn:results-g_n} arises from the multiplication of the analytically known divergent term with an $\mathcal{O}(\epsilon)$ expansion of the prefactor, see \eqref{eqn:vertex-diagrams}.}
\begin{equation}\label{eqn:g6}
  g_6 (u_{abcd}) = c_6 - 12\ln(2),\quad  c_6 = 5.57 \pm 0.05\,.
\end{equation}
In figure \ref{fig:Wvertextwogluonplot} we show the results for the two-gluon and vertex contributions for a continuously deformed kinematical configuration, generated as explained in app. \ref{app:generation-kinematics}, in order to illustrate, how the different contributions vary while their sum remains constant.\\

\begin{figure}[h]
\centering
 \includegraphics[width=1 \columnwidth]{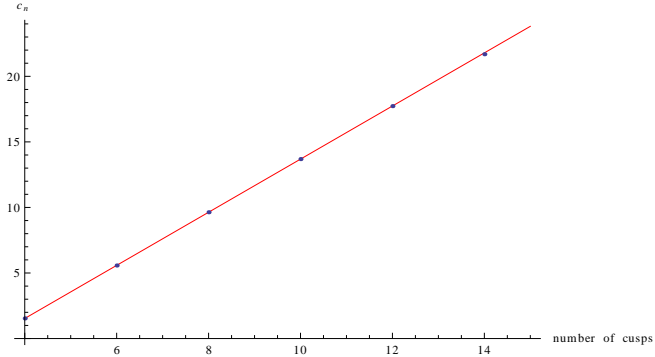}
\caption{This figure shows the dependence of the constant $c_n$ \eqref{eqn:n-dependence-constant} on the number of cusps $n$ of the Wilson loop.}
\label{fig:constant-dependence-on-number-of-cusps}
\end{figure}

~\\
{\bf Generalization to $n$ Cusps}\\~\\
It  turns out that also for $n>6$ the function of conformal cross ratios is just a constant, i.e.
\begin{equation}\label{eqn:results-g_n}
 g_n  (u_{abcd})  = c_n -  2 n \ln(2)\,.
\end{equation}
In figure \ref{fig:constant-dependence-on-number-of-cusps} we show the dependence of the numerical constant $c_n$ on the number of cusps $n$ up to $n=14$. Clearly, the constant depends linearly on the number of cusps $n$. It seems reasonable to assume that this dependence holds for all $n$, i.e.  
\begin{equation}\label{eqn:n-dependence-constant}
 c_n = a + b \cdot n\,,
\end{equation}
which is the line shown in fig. \ref{fig:constant-dependence-on-number-of-cusps} with the parameters\footnote{We determine the constants in \eqref{eqn:n-dependence-constant} from the results at $n=4$ and $n=6$, since here we have the smallest number of integrals and thus the best numerical result. At n-points we have to evaluate $2 {n \choose 3} + 2 {n \choose 4}$  two-gluon  \eqref{eqn:two-gluon-diagrams} and vertex integrals \eqref{eqn:vertex-finite-pieces}. 
}  
$a=6.6\pm0.1$,  $b = -2.028\pm 0.025$. 
Thus, we expect the Chern-Simons contribution to the $n$-point Wilson loop to be
\begin{align}\label{eqn:n-point-conjecture-CS} 
\langle W_n \rangle^{\text{CS}} &= - \frac{1}{4} \left( \frac{N}{k} \right)^2 \Bigg[2 \ln (2)\sum_{i=1}^n \frac{(-x_{i,i+2}^2 \tilde{\mu}^2  )^{2\epsilon}}{2 \epsilon} + g_n \Bigg]
\end{align}
where $g_n$ is the constant given by \eqref{eqn:results-g_n} and \eqref{eqn:n-dependence-constant}.

\section{ABJM theory}
In ABJM theory we have two gauge fields $A_\mu$, $\hat{A}_\mu$ and use the Wilson loop operator proposed in \cite{Drukker:2008zx}, which is a linear combination of two Wilson loops, each with one of the gauge fields, see also \cite{Henn:2010ps}. The one-loop contributions are identical to pure Chern Simons theory and cancel each other\footnote{They are also zero seperately, as explained above.}. Both two-loop contributions yield the same result and thus, it is sufficient to calculate the Wilson loop with the gauge field $A_\mu$. In addition to the Chern-Simons contributions \eqref{eqn:n-point-conjecture-CS} we have contributions of bosonic and fermionic matter fields which appear in the one-loop corrected gluon-propagator
\begin{equation}
 \langle W_n \rangle^{\text{ABJM}}_{\text{2-loop}} =  \langle W_n \rangle^{\text{CS}}_{\text{2-loop}} +  \langle W_n \rangle^{\text{matter}}_{\text{2-loop}}\,.
\end{equation}
The matter contribution is similar to the one in $\mathcal{N}=4$ SYM \cite{Brandhuber:2007yx}, since the one-loop corrected propagator calculated in $d$ dimensions \cite{Henn:2010ps} is\footnote{We drop the derivative term, it
would not contribute to the gauge-invariant Wilson loop  as explained in \cite{Henn:2010ps}.} 
\begin{align}
G_{\mu\nu}^{(1)}(x) 
&= -\frac{1}{N}\left(\frac{N}{k} \right)^2 \pi^{2-d} \Gamma \left( \frac{d}{2}-1 \right)^2   \frac{\eta_{\mu\nu}}{(-x^2)^{d-2}} 
\,,
\end{align}
which up to two small differences is  the tree level $\mathcal{N}=4$ SYM gluon propagator. The first difference is a trivial prefactor, and the second is that since we are at two loops, the power of $1/x^2$ is $1-2 \epsilon$ here,
as opposed to $1-\epsilon$ in the one-loop computation in $\mathcal{N}=4$ SYM.
Thus, it is clear that the results will be very similar to the expectation value of the Wilson loop in $\mathcal{N}=4$ SYM.

\begin{figure}[t]
\centering
\subfigure[]{\includegraphics[width=.31 \columnwidth]{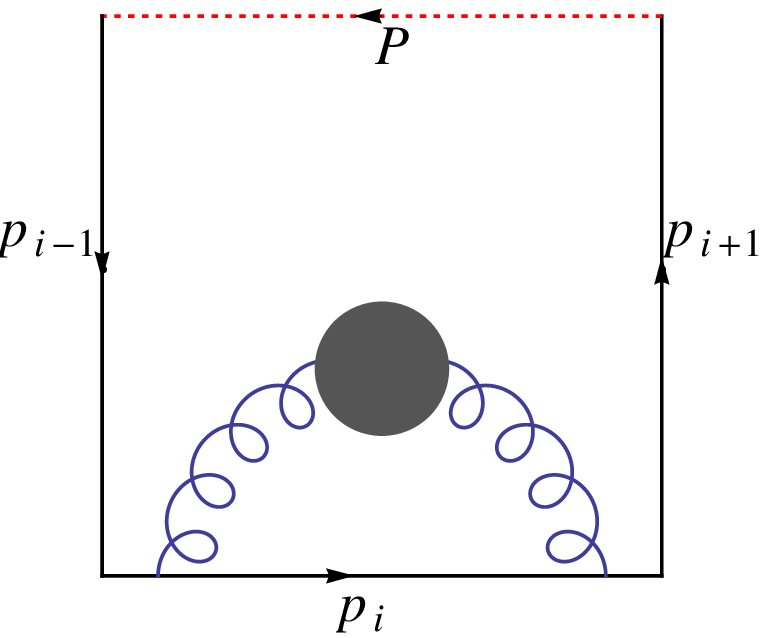}\label{fig:general-diagram-propagator-corrections-same-edge}}~~~~~
\subfigure[]{\includegraphics[width=.28 \columnwidth]{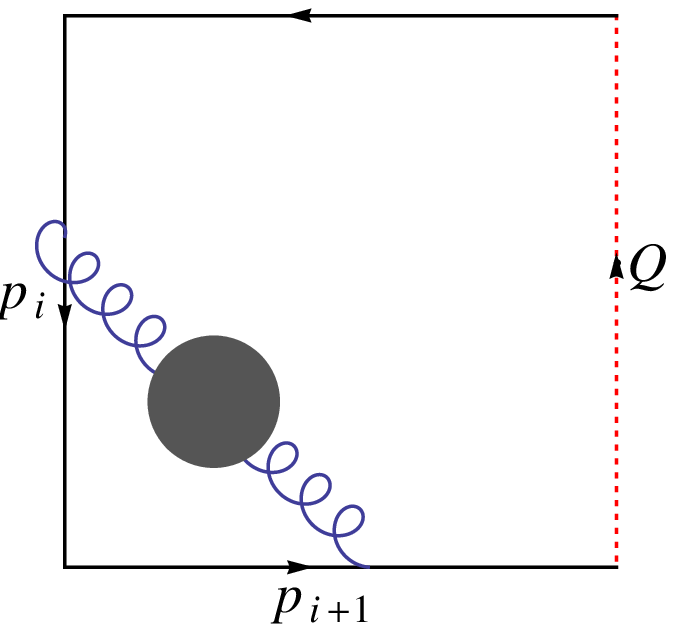}\label{fig:general-diagram-propagator-corrections-adjacent}}~~~~~
\subfigure[]{\includegraphics[width=.275 \columnwidth]{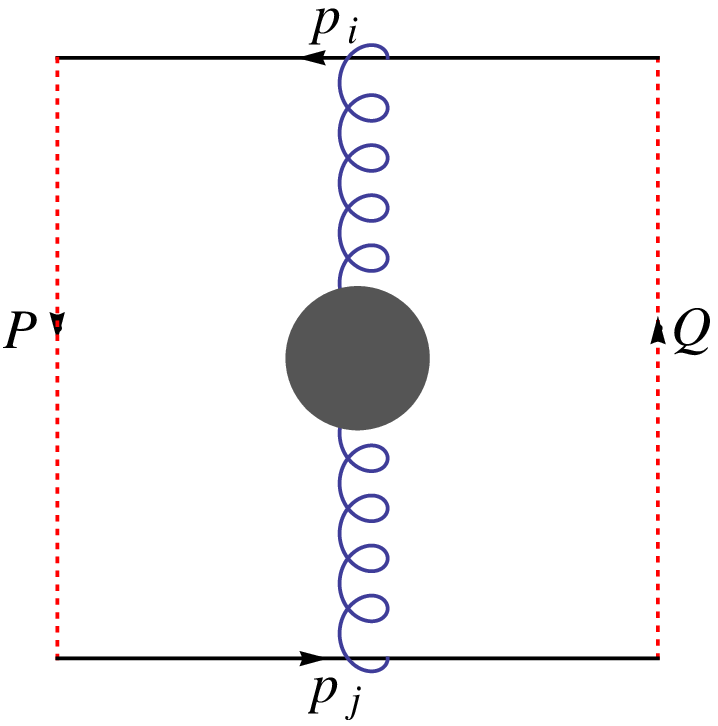}\label{fig:general-diagram-propagator-corrections-non-adjacent}}~~~~~~~~~~
\caption{The diagrams involving the one-loop corrected gluon propagator have the same structure as the 1-loop diagrams in $\mathcal{N}=4$ SYM. Dashed lines labelled with $P$ and $Q$ represent one or more light-like distances.
}
\label{fig:self-energy-diagrams}
\end{figure}

As in $\mathcal{N}=4$ SYM we have three classes of diagrams  shown in figure \ref{fig:self-energy-diagrams}. 
Diagram \ref{fig:general-diagram-propagator-corrections-same-edge} vanishes due to the light-likeness of the edges, 
whereas \ref{fig:general-diagram-propagator-corrections-adjacent} yields a divergent, and 
\ref{fig:general-diagram-propagator-corrections-non-adjacent} yields a finite contribution.

~\\
We have \cite{Henn:2010ps}
\begin{equation}\label{eqn:general-expression}\nn
\langle W_n \rangle^{\text{matter}}_{\text{2-loop}} 
= \left( \frac{N}{k}\right)^2 \left( (4 \pi e^{\gamma_E} )^{2 \epsilon} + \frac{\pi^2}{2}\epsilon^2 \right) \sum_{i>j} I_{ij}
\end{equation}
where 
\begin{align}
&I_{ij} 
= \frac{1}{2} \int_0^1 ds_i \int_0^1 ds_j  \\ \nn
&\frac{ x^2_{i,j+1} +  x^2_{i+1,j} -  x^2_{i,j} -  x^2_{i+1,j+1}}{(-x^2_{ij}\bar{s}_i\bar{s}_j - x^2_{i+1,j} s_i \bar{s}_j - x^2_{i,j+1}\bar{s}_i s_j - x^2_{i+1,j+1}s_i s_j)^{d-2}}\,.
\end{align}
There are $n$ divergent diagrams  $I_{i+1,i}$ of the type shown in fig. \ref{fig:general-diagram-propagator-corrections-adjacent}
\begin{equation}
\nn
I_{i+1,i} 
= -\frac{1}{8} \frac{(-x_{i,i+2}^2)^{2\epsilon}}{\epsilon^2} 
\end{equation}
and the finite diagrams $I_{ij}$ with $|i-j| \geq1$, see fig. \ref{fig:general-diagram-propagator-corrections-non-adjacent}, were solved in \cite{Brandhuber:2007yx} 
\begin{align}\nn
I_{ij} 
 = \frac{1}{2}\Big(&-\Li \left(1- a x_{i,j}^2 \right)-\Li \left(1- a x_{i+1,j+1}^2 \right) \\ \nn
& +\Li \left(1- a x_{i,j+1}^2 \right)+\Li \left(1- a x_{i+1,j}^2 \right)\Big)
\end{align}
where 
\begin{equation}\nn
a= \frac{x_{ij}^2 + x_{i+1,j+1}^2 -x_{i,j+1}^2-x_{i+1,j}^2}{x_{ij}^2 x_{i+1,j+1}^2 -x_{i,j+1}^2 x_{i+1,j}^2} \,.
\end{equation}
The sum over all finite diagrams $\sum_{i>j+1}^n I_{ij}= \mathcal{F}_n^{\text{WL}}$ is related to the well-known finite part of the BDS conjecture \cite{Bern:2005iz} via $\mathcal{F}_n^{\text{WL}}=\mathcal{F}_n^{\text{BDS}}- \frac{n}{4}\zeta(2)$, explicitly spelled out in \eqref{eqn:BDS-WL-n4-n6} for $n=4$ and $n=6$.

Then, the full matter part reads 
\begin{align}\nn
\langle W_n \rangle^{\text{matter}}_{\text{2-loop}} 
&= -\frac{1}{4} \left( \frac{N}{k}\right)^2 \Bigg[\sum_{i=1}^n \frac{(-x_{i,i+2}^2 \mu^2 4 \pi e^{\gamma_E})^{2\epsilon}}{2\epsilon^2} \\ \label{eqn:full-matter}
& \qquad \quad \qquad \quad -4 \mathcal{F}_n^{\text{WL}} + \frac{n}{4} \pi^2 \Bigg]\,.
\end{align}
where we have restored the regularisation scale $\mu$.
Taking into account the Chern-Simons result \eqref{eqn:n-point-conjecture-CS},  the full result in ABJM theory can be written as
\begin{align}\label{eqn:ABJM-result-Wn}\nn
\langle W_n \rangle^{\text{ABJM}}_{\text{2-loop}} &=  \left(\frac{N}{k}\right)^2 \Bigg[-\frac{1}{2}\sum_{i=1}^n \frac{(-x_{i,i+2}^2 \mu^2 8 \pi e^{\gamma_E})^{2\epsilon}}{(2\epsilon)^2} \\ 
& \qquad \qquad \qquad + \mathcal{F}_n^{\text{WL}} + r_n \Bigg]
\end{align}
where $r_n= -(\frac{n}{4} \pi^2 + c_n -2 n\ln (2) - 5 n  \ln^2(2))/4$ and $c_n$ is the numerical constant given by \eqref{eqn:n-dependence-constant}. 

Indeed, this is of the same form as the one-loop result for the Wilson loop in $\mathcal{N}=4$ SYM.

%

\section{Acknowledgements}
I would like to thank Marco S. Bianchi, Simon Caron-Huot, Wei-Ming Chen, Livia Ferro, Martin Heinze, Johannes Henn, Yu-tin Huang, Matias Leoni, Andrea Mauri, Silvia Penati, Jan Plefka, Andreas Rodigast, Radu Roiban, Alberto Santambrogio, Theodor Schuster and Sebastian Wuttke for useful discussions, correspondence and comments on the draft. Furthermore, I thank the organisers of the Mathematica summer school, the IGST workshop and conference for hospitality during my stay at the Perimeter Institute, where part of this work was carried out. In particular, I would like to thank Nikolay Gromov and Pedro Viera for their excellent lectures and tutorials on programming with Mathematica, which was extensively used in this project.

\appendix

\section{Two Loop Diagrams}\label{app:two-gluon-diagrams}
We use the same conventions as in \cite{Henn:2010ps}, i.e. the metric with $\eta_{\mu\nu}=\text{diag}(1,-1,-1)$ and we define an $n$-sided polygon by $n$ points $x_i$ ($i=1,..., n$), with the edge $i$ being the line connecting $x_i$ and $x_{i+1}$. Defining
\begin{equation}
p_i^{\mu}=x^{\mu}_{i+1}-x^{\mu}_i
\end{equation}
and parametrising the position $z^{\mu}_i$ on edge $i$ with the parameter $s_i \in [0,1]$ we have
\begin{equation} \label{eqn:z-parametrization}
z^{\mu}_i(s_i)= x^{\mu}_i + p^{\mu}_i s_i ,\quad z_{ij}^\mu := z_i^\mu - z_j^\mu
\end{equation}
Furthermore, we use the notations
\begin{equation}
\epsilon(p,q,r) = \epsilon_{\mu\nu\rho} p^{\mu} q^{\nu} r^{\rho} \qquad 
 \text{and} \qquad \bar{s}_i= 1- s_i.
\end{equation}
We use the Lagrangian of ABJM theory given in \cite{Henn:2010ps}.

\subsection{Two Gluon Diagrams}\label{app:two-gluon}
The contributions from the two-gluon diagrams are all finite and can be written as, see \cite{Henn:2010ps},
\begin{equation}\label{eqn:two-gluon-diagrams}
\langle W_n\rangle^{\text{two-gluon}}= \frac{1}{4} \left(\frac{N}{k}\right)^2 \sum_{i \geq j \geq k \geq l} \left( I_{ij,kl}+ I_{il,jk} \right)
\end{equation}
where
\begin{equation}\label{eqn:two-gluon-expr}
I_{ij,kl} = \int ds_{i,j,k,l} \frac{\epsilon(\dot{z}_i,\dot{z}_j,z_i-z_j)}{\left((z_i-z_j)^2\right)^{3/2}} \frac{\epsilon(\dot{z}_k,\dot{z}_l,z_k-z_l)}{\left((z_k-z_l)^2\right)^{3/2}}
\end{equation}
and the integration boundaries haven to be chosen according to the path ordering,  such that $z(s_i)>z(s_j)>z(s_k)>z(s_l)$. The diagram vanishes due to the epsilon tensor contractions if the gluon propagator of at least one of the index pairs connects the same or adjacent edges.

As a check of the numerics one can use the factorizing diagrams $I_{i+5,i+3,i+2,i}$, which are just a product of the analytically known one-loop diagrams \cite{Bianchi:2011rn}.

\subsection{Vertex Diagrams}\label{app:vertex}
The contribution from the vertex diagrams is calculated in the same way as in \cite{Henn:2010ps}, but generalized to $n$ points and we find\footnote{Details will be presented in \cite{wiegandt:2012phdthesis}.}
\begin{align}\label{eqn:vertex-diagrams}
\langle W_n \rangle^{\text{vertex}} 
&=  - \frac{1}{4} \left(\frac{N}{k}\right)^2 \left( \pi e^{\gamma_E}  \right)^{2 \epsilon} (1-2\epsilon) \\ \nn
&\quad \times  \frac{1}{4\pi} \sum_{i>j>k} \left(I_{ijk}^{(a)}+I_{ijk}^{(b)} \right)+ \mathcal{O}(\epsilon)\,.
\end{align}
The indices $i,j,k$ indicate the edges the gluon-propagators connect to and
\begin{align}\label{eqn:vertexIaandIb}\nn
I^{(a)}_{ijk} &= \int ds_{i,j,k},d[\beta]_3  \epsilon^{\alpha\beta\gamma}\epsilon_{\mu\alpha\sigma}\epsilon_{\nu\beta\lambda}\epsilon_{\rho\gamma\tau}  p_i^\mu p_j^\nu p_k^\rho \\ \nn 
& \quad \left[  \partial_j^\lambda  \Delta  \partial_i^\sigma  \partial_k^\tau \Delta+   \partial_i^\sigma \Delta \partial_j^\lambda  \partial_k^\tau \Delta+  \partial_k^\tau \Delta  \partial_i^\sigma  \partial_j^\lambda \Delta    \right] \Delta^{1-d} \\ \nn
I^{(b)}_{ijk} &=\int ds_{i,j,k},d[\beta]_3\epsilon^{\alpha\beta\gamma}\epsilon_{\mu\alpha\sigma}\epsilon_{\nu\beta\lambda}\epsilon_{\rho\gamma\tau}   p_i^\mu p_j^\nu p_k^\rho   \\ 
& \quad \left[\partial_i^\sigma \Delta \partial_j^\lambda \Delta \partial_k^\tau  \Delta \right] \Delta^{-d} (1-d)
\end{align}
where $\partial_i^\mu = \partial/\partial z_{i,\mu}$ and 
\begin{align}\nn
 \int d[\beta]_3 & = \int_0^1 d\beta_{i,j,k}\delta(\sum_m \beta_m-1)(\beta_i \beta_j \beta_k)^{d/2-2}, \\ \nn
\Delta & = - z_{ij}^2 \beta_i \beta_j- z_{ik}^2 \beta_i {\beta}_k- z_{kj}^2 \beta_k {\beta}_j\,.
\end{align}
Integrals with at least two propagators on the same edge (two identical indices $i,j,k$) vanish due to the antisymmetry of $I_{ijk}$ in the indices. Integrals $I^{(a)}_{ijk}$ of the type $i=k+2,j=k+1,k$ are divergent, see \cite{Henn:2010ps},  all other integrals are finite. 
The divergent integrals can be split up into a divergent and a finite piece
\begin{equation}\label{eqn:split-up-fin-div}
\langle W_n \rangle^{\text{vertex}} = \langle W_n \rangle^{\text{div}} + \langle W_n \rangle^{\text{finite}}\,.
\end{equation}
where the divergent piece can be evaluated analytically and reads
\begin{equation}
\langle W_n \rangle^{\text{div}} = - \left( \frac{N}{k} \right)^2 \left( \frac{\ln (2)}{2}  \sum_{i=1}^n \frac{(-x_{i,i+2}^2 \mu^2  \pi e^{\gamma_E})^{2\epsilon}}{2 \epsilon} \right)\,.
\end{equation}
The reamining finite piece is
\begin{align}\label{eqn:vertex-finite-pieces}
\langle W_n \rangle^{\text{finite}} &\\ \nn
=  - \frac{1}{4} \left(\frac{N}{k}\right)^2 & \left( \frac{1}{4\pi} \sum_{i>j>k} \left(I_{ijk}^{(a,f)}+I_{ijk}^{(b)} \right)- n \,2 \ln(2)\right)
\end{align}
where $I^{(a,f)},I^{(b)}$ are given by the following expressions for the case $i=k+2,j=k+1,k$. All other cases are treated purely numerically starting from \eqref{eqn:vertexIaandIb}, i.e. in particular $I^{(a,f)}_{ijk}=I^{(a)}_{ijk}$ in these cases.

For $i=3,j=2,k=1$, after solving two integrations and changing integration parameters to $\beta_i=x y, \beta_j= \bar{x}y, \beta_k=\bar{y}$ (where $\bar{x}=1-x, \bar{y}=1-y$), we have
\begin{align}\label{eqn:Iaf-expression-for-numerical-evaluation} \nn
 & I_{321}^{(a,f)} = -\frac{1}{2}\int_0^1 ds_j dx dy \frac{(x \bar{x} \bar{y})^{1/2}}{ e^\prime f^\prime} \\ \nn 
&\left(  \ln \left( e^\prime \frac{x_{13}^2}{x_{24}^2} + f^\prime -\frac{g}{x_{24}^2} \right)+ \ln \left( e^\prime + f^\prime \frac{x_{24}^2}{x_{13}^2}-\frac{g}{x_{13}^2} \right) \right)
 \\ \nn &+ \int_0^1 ds_j dx dy \frac{(x \bar{x} \bar{y})^{1/2}}{ e^\prime f^\prime} \left(1+ \ln(x\bar{x}\bar{y}) + \ln(e^\prime f^\prime)\right)
\end{align}
where the last term does not depend on the kinematical quantities and can be further integrated analytically and/or evaluated to high numerical precision. We used the abbreviations 
\begin{align} \nn
e  &= - x_{13}^2 e^\prime = -x_{13}^2 \bar{y}(x+s_j \bar{x}), \\ \nn
f  &= - x_{24}^2 f^\prime  = -x_{24}^2 x(\bar{y} +y \bar{s}_j \bar{x}) \\ \nn
g  &=  (x_{13}^2+x_{24}^2-x_{14}^2)x \bar{y}\,.
\end{align} Furthermore, after solving two integrations we find
\begin{align}\nn
I^{(b)}_{321} = 2  \int_0^1 dx dy ds_j (x \bar{y})^{3/2} \bar{x}^{1/2} \frac{(x_{14}^2-x_{13}^2-x_{24}^2)}{e^\prime f^\prime (e+f+g)}\,.
\end{align}
These are the expressions we use for the numerical evaluation of the finite parts of the divergent diagrams.

\subsection{Generation of kinematical configurations}\label{app:generation-kinematics}
A set of $n$ light-like ($p_i^2=0$) vectors $p_i^\mu$ satisfying momentum conservation $\sum p_i^\mu=0$ can easily be generated by choosing the $p^0_i$ components of $n-3$ vectors and the angle $\theta_i$ between $p_i^1 =p_i^0 \cos \theta_i $, $p_i^2 =p_i^0 \sin \theta_i$. The remaining components are then fixed. For $n$ even it is possible to choose configurations, where
all non-light-like distances $x_{ij}^2$ are space-like. For the numerical evaluation we make use of this type of configurations, such that all integrals are real.

The configurations used for the results shown in fig. 
\ref{fig:Wvertextwogluonplot} are obtained by continuously deforming two angles $\theta_i$, leading to conformally non-equivalent kinematical configurations. We use the angles
\begin{align}
 \theta_i(a):= \pi \left\{\frac{16}{9}a,\frac{13 }{9},\frac{5}{3}a,\frac{13  }{8},\frac{19 }{14},1\right\}
\end{align}
and choose $p_i^{\mu=0}=\{1,-3,4\}$, the remaining components are then fixed. The parameter $a$ is chosen between $a=1$ and $a=1.2$ in steps of $0.01$.

 \vfill


\begin{thebibliography}{10}
\ifx\href\asklfhas\newcommand{\href}[2]{#2}\fi
\ifx\arxivref\asklfhas\newcommand{\arxivref}[2]{\href{http://arxiv.org/abs/#1}%
{#2}}\fi
\ifx\doiref\asklfhas\newcommand{\doiref}[2]{\href{http://dx.doi.org/#1}{#2}}\fi
\raggedright
\small
\parskip 0pt

\bibitem{Aharony:2008ug}
O.~Aharony, O.~Bergman, D.~L.~Jafferis and J.~Maldacena,
\textit{``{{$\mathcal{N}=\mathord{}$6} superconformal Chern-Simons-matter
  theories, M2-branes and their gravity duals}''},
\textsf{\doiref{10.1088/1126-6708/2008/10/091}{JHEP~0810,~091~(2008)}},
\texttt{\arxivref{0806.1218}{arxiv:0806.1218}}.

\bibitem{Alday:2007hr}
L.~F.~Alday and J.~M.~Maldacena,
\textit{``{Gluon scattering amplitudes at strong coupling}''},
\textsf{\doiref{10.1088/1126-6708/2007/06/064}{JHEP~0706,~064~(2007)}},
\texttt{\arxivref{0705.0303}{arxiv:0705.0303}}.

\bibitem{Drummond:2007aua}
J.~M.~Drummond, G.~P.~Korchemsky and E.~Sokatchev,
\textit{``{Conformal properties of four-gluon planar amplitudes and Wilson
  loops}''},
\textsf{\doiref{10.1016/j.nuclphysb.2007.11.041}{Nucl.~Phys.~B795,~385~(2008)}%
},
\texttt{\arxivref{0707.0243}{arxiv:0707.0243}}.

\bibitem{Brandhuber:2007yx}
A.~Brandhuber, P.~Heslop and G.~Travaglini,
\textit{``{MHV Amplitudes in {$\mathcal{N}=\mathord{}$4} Super Yang--Mills and
  Wilson Loops}''},
\textsf{\doiref{10.1016/j.nuclphysb.2007.11.002}{Nucl.~Phys.~B794,~231~(2008)}%
},
\texttt{\arxivref{0707.1153}{arxiv:0707.1153}}.

\bibitem{Drummond:2007cf}
J.~M.~Drummond, J.~Henn, G.~P.~Korchemsky and E.~Sokatchev,
\textit{``{On planar gluon amplitudes/Wilson loops duality}''},
\textsf{\doiref{10.1016/j.nuclphysb.2007.11.007}{Nucl.~Phys.~B795,~52~(2008)}},
\texttt{\arxivref{0709.2368}{arxiv:0709.2368}}.

\bibitem{Drummond:2008vq}
J.~M.~Drummond, J.~Henn, G.~P.~Korchemsky and E.~Sokatchev,
\textit{``{Dual superconformal symmetry of scattering amplitudes in
  {$\mathcal{N}=\mathord{}$4} super-Yang--Mills theory}''},
\textsf{\doiref{10.1016/j.nuclphysb.2009.11.022}{Nucl.~Phys.~B828,~317~(2010)}%
},
\texttt{\arxivref{0807.1095}{arxiv:0807.1095}}.

\bibitem{Drummond:2009fd}
J.~M.~Drummond, J.~M.~Henn and J.~Plefka,
\textit{``{Yangian symmetry of scattering amplitudes in
  {$\mathcal{N}=\mathord{}$4} super Yang-Mills theory}''},
\textsf{\doiref{10.1088/1126-6708/2009/05/046}{JHEP~0905,~046~(2009)}},
\texttt{\arxivref{0902.2987}{arxiv:0902.2987}}.

\bibitem{Alday:2008yw}
L.~F.~Alday and R.~Roiban,
\textit{``{Scattering Amplitudes, Wilson Loops and the String/Gauge Theory
  Correspondence}''},
\textsf{\doiref{10.1016/j.physrep.2008.08.002}{Phys.~Rept.~468,~153~(2008)}},
\texttt{\arxivref{0807.1889}{arxiv:0807.1889}}.

\bibitem{Henn:2009bd}
J.~Henn,
\textit{``{Duality between Wilson loops and gluon amplitudes}''},
\textsf{\doiref{10.1002/prop.200900048}{Fortsch.Phys.~57,~729~(2009)}},
\texttt{\arxivref{0903.0522}{arxiv:0903.0522}},
Based on the author's Ph.D. thesis at the University Lyon I (France and
  prepared at LAPTH, Annecy-le-Vieux (France).

\bibitem{Beisert:2010jr}
N.~Beisert, C.~Ahn, L.~F.~Alday, Z.~Bajnok, J.~M.~Drummond et~al.,
\textit{``{Review of AdS/CFT Integrability: An Overview}''},
\texttt{\arxivref{1012.3982}{arxiv:1012.3982}},
Long author list - awaiting processing.

\bibitem{Drummond:2007bm}
J.~M.~Drummond, J.~Henn, G.~P.~Korchemsky and E.~Sokatchev,
\textit{``{The hexagon Wilson loop and the BDS ansatz for the six-gluon
  amplitude}''},
\textsf{\doiref{10.1016/j.physletb.2008.03.032}{Phys.~Lett.~B662,~456~(2008)}},
\texttt{\arxivref{0712.4138}{arxiv:0712.4138}}.

\bibitem{Bern:2008ap}
Z.~Bern, L.~J.~Dixon, D.~A.~Kosower, R.~Roiban, M.~Spradlin, C.~Vergu and
  A.~Volovich,
\textit{``{The Two-Loop Six-Gluon MHV Amplitude in Maximally Supersymmetric
  Yang-Mills Theory}''},
\textsf{\doiref{10.1103/PhysRevD.78.045007}{Phys.~Rev.~D78,~045007~(2008)}},
\texttt{\arxivref{0803.1465}{arxiv:0803.1465}}.

\bibitem{Bern:2005iz}
Z.~Bern, L.~J.~Dixon and V.~A.~Smirnov,
\textit{``Iteration of planar amplitudes in maximally supersymmetric Yang-Mills
  theory at three loops and beyond''},
\textsf{\doiref{10.1103/PhysRevD.72.085001}{Phys.~Rev.~D72,~085001~(2005)}},
\texttt{\arxivref{hep-th/0505205}{hep-th/0505205}}.

\bibitem{CaronHuot:2010ek}
S.~Caron-Huot,
\textit{``{Notes on the scattering amplitude / Wilson loop duality}''},
\textsf{\doiref{10.1007/JHEP07(2011)058}{JHEP~1107,~058~(2011)}},
\texttt{\arxivref{1010.1167}{arxiv:1010.1167}}.

\bibitem{Mason:2010yk}
L.~Mason and D.~Skinner,
\textit{``{The Complete Planar S-matrix of N=4 SYM as a Wilson Loop in Twistor
  Space}''},
\textsf{\doiref{10.1007/JHEP12(2010)018}{JHEP~1012,~018~(2010)}},
\texttt{\arxivref{1009.2225}{arxiv:1009.2225}}.

\bibitem{Alday:2010zy}
L.~F.~Alday, B.~Eden, G.~P.~Korchemsky, J.~Maldacena and E.~Sokatchev,
\textit{``{From correlation functions to Wilson loops}''},
\texttt{\arxivref{1007.3243}{arxiv:1007.3243}}.

\bibitem{Eden:2010ce}
B.~Eden, G.~P.~Korchemsky and E.~Sokatchev,
\textit{``{More on the duality correlators/amplitudes}''},
\texttt{\arxivref{1009.2488}{arxiv:1009.2488}}.

\bibitem{Eden:2011yp}
B.~Eden, P.~Heslop, G.~P.~Korchemsky and E.~Sokatchev,
\textit{``{The super-correlator/super-amplitude duality: Part I}''},
\texttt{\arxivref{1103.3714}{arxiv:1103.3714}},
* Temporary entry *.

\bibitem{Eden:2011ku}
B.~Eden, P.~Heslop, G.~P.~Korchemsky and E.~Sokatchev,
\textit{``{The super-correlator/super-amplitude duality: Part II}''},
\texttt{\arxivref{1103.4353}{arxiv:1103.4353}},
* Temporary entry *.

\bibitem{Adamo:2011dq}
T.~Adamo, M.~Bullimore, L.~Mason and D.~Skinner,
\textit{``{A Proof of the Supersymmetric Correlation Function / Wilson Loop
  Correspondence}''},
\textsf{\doiref{10.1007/JHEP08(2011)076}{JHEP~1108,~076~(2011)}},
\texttt{\arxivref{1103.4119}{arxiv:1103.4119}},
* Temporary entry *.

\bibitem{Berkovits:2008ic}
N.~Berkovits and J.~Maldacena,
\textit{``{Fermionic T-Duality, Dual Superconformal Symmetry, and the
  Amplitude/Wilson Loop Connection}''},
\textsf{\doiref{10.1088/1126-6708/2008/09/062}{JHEP~0809,~062~(2008)}},
\texttt{\arxivref{0807.3196}{arxiv:0807.3196}}.

\bibitem{Beisert:2008iq}
N.~Beisert, R.~Ricci, A.~A.~Tseytlin and M.~Wolf,
\textit{``{Dual Superconformal Symmetry from $AdS_5 \times S^5$ Superstring
  Integrability}''},
\textsf{\doiref{10.1103/PhysRevD.78.126004}{Phys.~Rev.~D78,~126004~(2008)}},
\texttt{\arxivref{0807.3228}{arxiv:0807.3228}}.

\bibitem{Grassi:2009yj}
P.~A.~Grassi, D.~Sorokin and L.~Wulff,
\textit{``{Simplifying superstring and D-brane actions in $AdS_4 x CP^3$
  superbackground}''},
\textsf{\doiref{10.1088/1126-6708/2009/08/060}{JHEP~0908,~060~(2009)}},
\texttt{\arxivref{0903.5407}{arxiv:0903.5407}}.

\bibitem{Adam:2010hh}
I.~Adam, A.~Dekel and Y.~Oz,
\textit{``{On the fermionic T-duality of the $AdS_4 x CP^3$ sigma-model}''},
\textsf{\doiref{10.1007/JHEP10(2010)110}{JHEP~1010,~110~(2010)}},
\texttt{\arxivref{1008.0649}{arxiv:1008.0649}}.

\bibitem{Adam:2009kt}
I.~Adam, A.~Dekel and Y.~Oz,
\textit{``{On Integrable Backgrounds Self-dual under Fermionic T-duality}''},
\textsf{\doiref{10.1088/1126-6708/2009/04/120}{JHEP~0904,~120~(2009)}},
\texttt{\arxivref{0902.3805}{arxiv:0902.3805}}.

\bibitem{Dekel:2011qw}
A.~Dekel and Y.~Oz,
\textit{``{Self-Duality of Green-Schwarz Sigma-Models}''},
\textsf{\doiref{10.1007/JHEP03(2011)117}{JHEP~1103,~117~(2011)}},
\texttt{\arxivref{1101.0400}{arxiv:1101.0400}}.

\bibitem{Bakhmatov:2010fp}
I.~Bakhmatov,
\textit{``{On $AdS_4$ x $CP^3$ T-duality}''},
\textsf{\doiref{10.1016/j.nuclphysb.2011.01.020}{Nucl.Phys.~B847,~38~(2011)}},
\texttt{\arxivref{1011.0985}{arxiv:1011.0985}}.

\bibitem{Bakhmatov:2011aa}
I.~Bakhmatov, E.~O.~Colgain and H.~Yavartanoo,
\textit{``{Fermionic T-duality in the pp-wave limit}''},
\texttt{\arxivref{1109.1052}{arxiv:1109.1052}},
* Temporary entry *.

\bibitem{Bargheer:2010hn}
T.~Bargheer, F.~Loebbert and C.~Meneghelli,
\textit{``{Symmetries of Tree-level Scattering Amplitudes in N=6 Superconformal
  Chern-Simons Theory}''},
\texttt{\arxivref{1003.6120}{arxiv:1003.6120}}.

\bibitem{Lee:2010du}
S.~Lee,
\textit{``{Yangian Invariant Scattering Amplitudes in Supersymmetric
  Chern-Simons Theory}''},
\textsf{\doiref{10.1103/PhysRevLett.105.151603}{Phys.Rev.Lett.~105,~151603~(20%
10)}},
\texttt{\arxivref{1007.4772}{arxiv:1007.4772}}.

\bibitem{Huang:2010qy}
Y.-t.~Huang and A.~E.~Lipstein,
\textit{``{Dual Superconformal Symmetry of N=6 Chern-Simons Theory}''},
\textsf{\doiref{10.1007/JHEP11(2010)076}{JHEP~1011,~076~(2010)}},
\texttt{\arxivref{1008.0041}{arxiv:1008.0041}}.

\bibitem{Gang:2010gy}
D.~Gang, Y.-t.~Huang, E.~Koh, S.~Lee and A.~E.~Lipstein,
\textit{``{Tree-level Recursion Relation and Dual Superconformal Symmetry of
  the ABJM Theory}''},
\textsf{\doiref{10.1007/JHEP03(2011)116}{JHEP~1103,~116~(2011)}},
\texttt{\arxivref{1012.5032}{arxiv:1012.5032}}.

\bibitem{Lipstein:2011ej}
A.~E.~Lipstein,
\textit{``{Integrability of N = 6 Chern-Simons Theory}''},
\texttt{\arxivref{1105.3231}{arxiv:1105.3231}}.

\bibitem{Agarwal:2008pu}
A.~Agarwal, N.~Beisert and T.~McLoughlin,
\textit{``{Scattering in Mass-Deformed $N>=4$ Chern-Simons Models}''},
\textsf{\doiref{10.1088/1126-6708/2009/06/045}{JHEP~0906,~045~(2009)}},
\texttt{\arxivref{0812.3367}{arxiv:0812.3367}}.

\bibitem{Henn:2010ps}
J.~M.~Henn, J.~Plefka and K.~Wiegandt,
\textit{``{Light-like polygonal Wilson loops in 3d Chern-Simons and ABJM
  theory}''},
\textsf{\doiref{10.1007/JHEP08(2010)032}{JHEP~1008,~032~(2010)}},
\texttt{\arxivref{1004.0226}{arxiv:1004.0226}}.

\bibitem{Bianchi:2011rn}
M.~S.~Bianchi, M.~Leoni, A.~Mauri, S.~Penati, C.~A.~Ratti et~al.,
\textit{``{From Correlators to Wilson Loops in Chern-Simons Matter
  Theories}''},
\textsf{\doiref{10.1007/JHEP06(2011)118}{JHEP~1106,~118~(2011)}},
\texttt{\arxivref{1103.3675}{arxiv:1103.3675}},
* Temporary entry *.

\bibitem{Chen:2011vv}
W.-M.~Chen and Y.-t.~Huang,
\textit{``{Dualities for Loop Amplitudes of N=6 Chern-Simons Matter Theory}''},
\texttt{\arxivref{1107.2710}{arxiv:1107.2710}},
* Temporary entry *.

\bibitem{Bianchi:2011dg}
M.~S.~Bianchi, M.~Leoni, A.~Mauri, S.~Penati and A.~Santambrogio,
\textit{``{Scattering Amplitudes/Wilson Loop Duality In ABJM Theory}''},
\texttt{\arxivref{1107.3139}{arxiv:1107.3139}}.

\bibitem{Bianchi:2011fc}
M.~S.~Bianchi, M.~Leoni, A.~Mauri, S.~Penati and A.~Santambrogio,
\textit{``{Scattering in ABJ theories}''},
\texttt{\arxivref{1110.0738}{arxiv:1110.0738}},
* Temporary entry *.

\bibitem{Chen:1992ee}
W.~Chen, G.~W.~Semenoff and Y.-S.~Wu,
\textit{``{Two loop analysis of non Abelian Chern-Simons theory}''},
\textsf{\doiref{10.1103/PhysRevD.46.5521}{Phys.~Rev.~D46,~5521~(1992)}},
\texttt{\arxivref{hep-th/9209005}{hep-th/9209005}}.

\bibitem{Drukker:2008zx}
N.~Drukker, J.~Plefka and D.~Young,
\textit{``{Wilson loops in 3-dimensional N=6 supersymmetric Chern- Simons
  Theory and their string theory duals}''},
\textsf{\doiref{10.1088/1126-6708/2008/11/019}{JHEP~0811,~019~(2008)}},
\texttt{\arxivref{0809.2787}{arxiv:0809.2787}}.

\bibitem{wiegandt:2012phdthesis}
K.~Wiegandt,
\textit{``{PhD thesis, Humboldt University}''}.

\end{thebibliography}

\end{document}